# Extreme near-field heat transfer between silica surfaces

Ali Rajabpour[1,2,*], Julien El Hajj[2], M. Gómez Viloria[3], Riccardo Messina[3], Philippe Ben-Abdallah[3], Yangyu Guo[4] and Samy Merabia[2,*]

[1]Mechanical Engineering Department, Imam Khomeini International University, Qazvin 34148-96818, Iran
[2]Institut Lumière Matière, Université Claude Bernard Lyon 1-CNRS, Université de Lyon, Villeurbanne 69622, France
[3]Laboratoire Charles Fabry, UMR 8501, Institut d'Optique ,Université Paris-Saclay, 91127 Palaiseau Cedex, France
[4]School of Energy Science and Engineering, Harbin Institute of Technology, Harbin 150001, China

**Abstract**

Despite recent experiments exhibiting an impressive enhancement in radiative heat flux between parallel planar silica surfaces with gap sizes of about 10 nm, the exploration of sub-nanometric gap distances remains unexplored. In this work, by employing non-equilibrium molecular dynamics (NEMD) simulations, we study the heat transfer between two $SiO_2$ plates in both their amorphous and crystalline forms. When the gap size is 2 nm, we find that the heat transfer coefficient experiences a substantial ~30-fold increase compared to the experimental value at the gap size of 10 nm confirming the dependence on the distance inversely quadratic as predicted by the fluctuational electrodynamics (FE) theory. Comparative analysis between NEMD and FE reveals a generally good agreement, particularly for amorphous silica. Spectral heat transfer analysis demonstrates the profound influence of gap size on heat transfer, with peaks corresponding to the resonances of dielectric function. Deviations from fluctuational electrodynamics theory at smaller gap sizes are interpreted in the context of acoustic phonon tunneling and the effects of a gradient of permittivity close to the surfaces.

The classical theory of thermal radiation, based on the black body radiation model, assumes characteristic lengths larger than the thermal wavelength ($\lambda_{th}$ = 2898/T [μm]). In the near-field regime, i.e. the scenario in which these lengths become comparable, wave characteristics must be considered [1–4]. This leads to tunneling evanescent modes, potentially resulting in radiative heat transfer several orders of magnitude higher than predicted by Stefan-Boltzmann's law [5–7]. Investigating this regime is important for applications including scanning thermal microscopy, thermally assisted magnetic recording, nano-lithography, near-field thermophotovoltaics, and non-contact friction [8–17].

---

[*] Corresponding authors, Email: rajabpour@eng.ikiu.ac.ir (A. Rajabpour) and samy.merabia@univ-lyon1.fr (S. Merabia)

Extreme near-field heat transfer between closely spaced bodies (less than 10 nm) involves photons, phonons, and electrons [18–30]. The fluctuational electrodynamics (FE) theory is necessary for analyzing near-field radiative heat transfer involving photons wherein the conventional blackbody theorem fails [31–39]. In this context, the work by Mulet et al. demonstrated a substantial enhancement in radiative heat transfer at a subwavelength scale, mainly attributed to resonant surface waves [4]. Phonons, both optical and acoustic may contribute to near-field heat transfer, and recent studies have highlighted the role of phonon tunneling, especially across vacuum gaps [26,27,40–42]. Electronic contributions become significant only at angstrom distances, dominating heat transfer at sub-nanometric distances [19,20,26,43].

The role that phonons play in heat transfer between two bodies is usually taken into account by using dielectric permittivity in calculating the radiation contribution but excludes acoustic phonons via tunneling which can be another channel of extreme near-field heat transfer. Altfeder et al., using ultrahigh vacuum inelastic scanning tunneling microscopy, revealed that thermal vibrations from the tip atoms effectively transmit their energy to the sample surface across a narrow vacuum gap leading to phonon tunneling enhanced heat transfer [9]. Prunnila et al. elucidated how acoustic phonons can directly tunnel through a vacuum between piezoelectric materials, facilitating energy transmission and heat conduction between bodies separated by a vacuum gap through an electrostatic coupling mechanism [44]. Recently, Guo et al. focused on atomistic simulation of phonon heat transport across metallic and polar vacuum nanogaps [28,29,45]. Their results showed significant heat transport below a gap size of 1 nm, with lattice anharmonicity contributing to a substantial portion of phonon tunneling. They showed that non-local dielectric response can influence heat transport in the extreme near-field regime between polar MgO materials. However, it was shown that local and non-local models yield the same heat fluxes for gaps larger than 2 nm between metallic surfaces [46].

Silicon dioxide ($SiO_2$) plays a significant role in extreme near-field heat transfer due to its dielectric properties, spacer material, compatibility with various materials, and potential for thermal insulation [35,36,43,47–49]. Experimental and computational studies have explored heat transfer between two $SiO_2$ bodies separated by micrometric and nanometric gaps [35,50,51]. These studies have reported substantial enhancements in radiative heat transfer, thermal coupling, and rectification, providing insights into the complex mechanisms governing extreme near-field heat transfer [37,52,53]. Notably, recent experiments by Garcia et al. and Fiorino et al. achieved more than a 1000-fold enhancement in radiative heat flux between parallel planar silica surfaces with gaps as small as ∼20 nm [47,54]. Moreover, Salihoglu et al. reported an experimental study of extreme near-field heat transfer between two parallel quartz plates with a distance of less than 10 nm and observed 18,000 enhancement in the heat transfer coefficient [55].

Despite these recent advancements, the range of gap distances explored in experimental studies on parallel $SiO_2$ plates in the extreme near-field heat transfer regime is generally above 10 nm, leaving the investigation of smaller gap distances unexplored. This study addresses this gap by employing non-equilibrium molecular dynamics simulations to characterize heat transfer for $SiO_2$ plates in both their amorphous and crystalline forms. The results are compared with the FE theory and highlight the specific frequencies contributing the most to heat transfer. Furthermore, we compare the NEMD results with a refined FE approach that incorporates non-local effects through the frequency and wave-vector-dependent dielectric function for crystalline silica.

To investigate near-field heat transfer, two silica bodies separated by a gap distance $d$ along the $z$-direction are considered, as shown in Figure 1. Atomic positions of both amorphous and crystalline forms have been generated via the VMD package [56]. Amorphous silica (glass) has a disordered structure, while crystalline silica has an $\alpha$-quartz structure with a cubic phase. Each silica slab has a length of approximately 10 nm and the gap between the two bodies is created by shifting the atomic positions to keep a net charge of zero. The lateral dimensions in the $x$ and $y$ directions are equal to about 5 nm, and periodic boundary conditions are applied to minimize size effects in these directions.

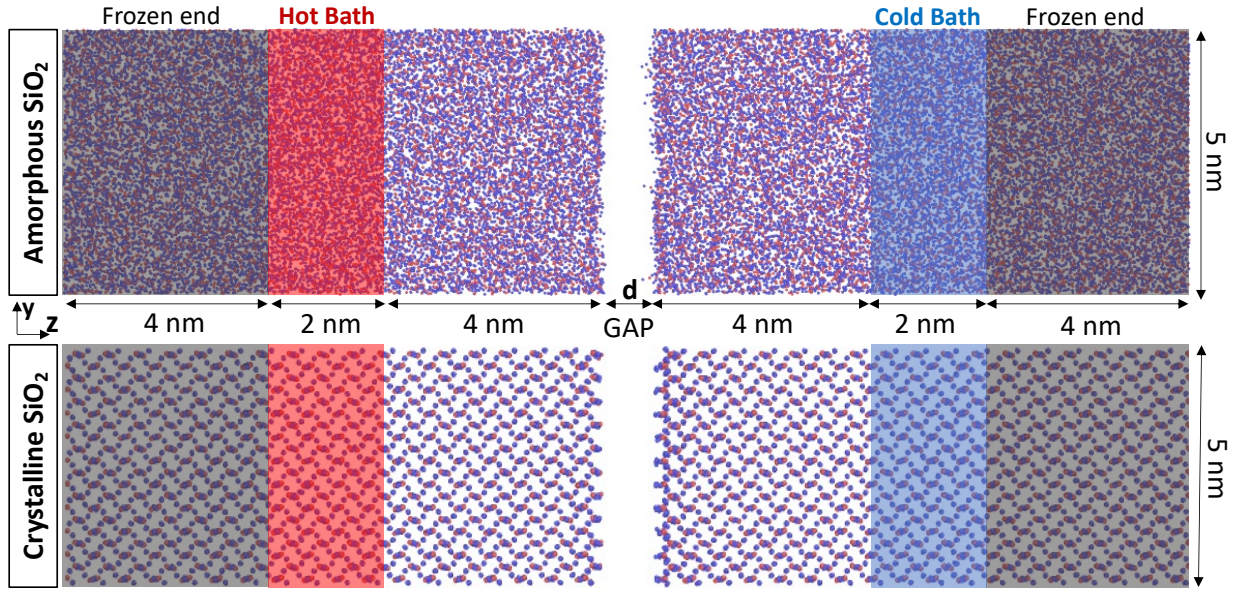

Fig. 1: Atomic structures of amorphous and crystalline SiO$_2$ and the NEMD setup to calculate the heat transfer through the gap having a distance *d*

The LAMMPS package is utilized to solve Newton's equations of motion in the molecular dynamics simulation method [57]. The long-range Beest Kramer van Santen (BKS) potential is employed to describe atomic interactions in silica [58,59]. This potential has been widely used to investigate the thermal properties of silica [50,60–62]. The partial charges of silicon and oxygen atoms are set to 2.4 and -1.2, respectively. An additional 6-24 Lennard-Jones potential with parameters reported in Ref. [63] is used in the amorphous form to avoid atoms getting very close together. The particle-particle particle-mesh (PPPM) method is employed to calculate long-range interactions in the system [57,64].

To calculate the heat transfer from non-equilibrium molecular dynamics, both boundaries of the system (with a thickness of ~ 4 nm) are frozen to prevent heat transfer through the periodic boundary condition along the *z*-direction as depicted in Figure 1. The hot and cold baths are created by two Nose-Hoover thermostats [65,66] with a size of 2 nm. A 4 nm distance is maintained between the baths and the gap region to ensure that heat transfer in the central region is not affected by thermostats.

To preserve the atomic structure of crystalline silica, the free surface (~ 2 Å) near the gap is tethered in the *z*-direction by a harmonic spring with a force constant of 80 N/m, while for amorphous silica, this tethering is unnecessary except for gap sizes less than 5 Å. The temperature difference applied to the system is $\Delta T$=100 K to reduce statistical uncertainties in the NEMD method.

The heat transfer coefficient $h$ between the two silica bodies is computed as $h = q/A\Delta T$, where $q$ is the net heat flow through the gap. To calculate the spectral heat transfer in terms of frequency, the method introduced by Sääskilahti et al. is employed [67–69]. Note that Guo et al. have provided more details about calculating spectral heat flux by molecular dynamics [28,45]. The spectral heat transfer coefficient is calculated using the equation $h(\omega) = q(\omega)/A\Delta T$, where $\omega$ is the angular frequency and $q(\omega)$ is the spectral heat flux which is computed as follows:

$$q(\omega) = 2\,\text{Re}\left[\sum_{i\in I}\int_0^\infty \langle \boldsymbol{F}_i(t)\cdot\boldsymbol{v}_i(\boldsymbol{0})\rangle e^{-j\omega t}dt\right] \quad (1)$$

$$q = \int_0^\infty q(\omega)\frac{d\omega}{2\pi} \quad (2)$$

Here, Re means the real part, $\boldsymbol{F}_i$ is the net force applied to each atom at one side of the gap ($I$) only due to the atoms on the other side of the gap. $\boldsymbol{v}$ is the atomic velocity, and $\langle \cdots \rangle$ denotes an ensemble average.

To apply quantum corrections, $q(\omega)$ is multiplied by a factor of $\hbar\omega\,\partial f_{\text{BE}}/\partial T$, where $f_{\text{BE}}$ is the Bose-Einstein distribution function and $\hbar$ is the Planck constant [28].

In parallel, we compare NEMD results with fluctuational electrodynamics (FE) theory [4,7] allowing us to calculate radiative heat transfer between two semi-infinite parallel plates with an extremely close distance in vacuum conditions. Heat transfer between two bodies 1 and 2 with a gap size of $d$ is the sum of evanescent waves and propagative waves contributions as $q = q_{\text{evanescent}} + q_{\text{propagative}}$. In this case, the heat transfer coefficient between two bodies 1 and 2 is approximated as follows [4]:

$$h(\omega) = \frac{1}{\pi d^2}\frac{\varepsilon_1''}{|1+\varepsilon_1|^2}\frac{\varepsilon_2''}{|1+\varepsilon_2|^2}k_B\left(\frac{\hbar\omega}{k_BT}\right)^2\frac{\exp(\hbar\omega/k_BT)}{[\exp(\hbar\omega/k_BT)-1]^2} \quad (3)$$

Here, $\varepsilon_i(\omega) = \varepsilon'_i(\omega) + j\,\varepsilon''_i(\omega)$ denotes the bulk complex permittivity of the two bodies ($i$ = 1,2), $T$ is the mean temperature of the system and $k_B$ is the Boltzmann's constant. From the FE approach, it is clear that the heat transfer coefficient is proportional to $d^{-2}$. To calculate the dielectric function of silica, equilibrium molecular dynamics simulations have been exploited based on the fluctuation-dissipation theorem [45,70,71]. Thus, the dielectric function of the bulk silica is derived from a Green-Kubo relation as follows:

$$\varepsilon(\omega) = 1 + \frac{V}{3\varepsilon_0 k_B T}[\langle \boldsymbol{P}(0).\boldsymbol{P}(0)\rangle + j\omega \int_0^\infty \langle \boldsymbol{P}(t).\boldsymbol{P}(0)\rangle e^{-j\omega t}dt\,] \quad (4)$$

where $V$ is the volume of the bulk system and $\varepsilon_0$ is the permittivity of vacuum. $\boldsymbol{P}(t) = (\sum_i e_i \boldsymbol{r}_i)/V$ where $\boldsymbol{r}_i$ is the atomic displacement vector of particle $i$ and $e_i$ is the corresponding partial charge.

It is possible to take into account the non-local behavior of the dielectric function (i.e. its dependence on the wavevector $\boldsymbol{k}$), which turns out to be relevant at short separation distances, by computing

$$\varepsilon(\omega, \boldsymbol{k}) = 1 + \frac{V}{3\varepsilon_0 k_B T}[\langle \boldsymbol{P}^*(0,\boldsymbol{k}).\boldsymbol{P}(0,\boldsymbol{k})\rangle + j\omega \int_0^\infty \langle \boldsymbol{P}^*(t,\boldsymbol{k}).\boldsymbol{P}(0,\boldsymbol{k})\rangle e^{-j\omega t}dt\,] \quad (5)$$

where * is the complex conjugate symbol and the polarization vector is calculated as a function of the wave vector and time in the form $\boldsymbol{P}(t) = (\sum_i e_i \boldsymbol{r}_i(t) e^{-j\boldsymbol{k}.\boldsymbol{r}_{0,i}})/V$ which is the spatial Fourier transform of the product of the partial charge ($e_i$) to the equilibrium positions ($\boldsymbol{r}_{0,i}$) of the atoms in the unit cell. It should be noted that in the calculation of dielectric functions performed in this study, the electronic contribution is not considered to make a fair comparison between the results of fluctuational electrodynamics and molecular dynamics simulations, which considers only vibrational degrees of freedom.

The non-equilibrium molecular dynamics (NEMD) simulation method was employed to calculate heat transfer for gap sizes ranging from ~0.5 nm to 2 nm. In Figure 2a, the heat transfer coefficient for $SiO_2$-$SiO_2$ gap is observed to be higher than that of non-polar Au-Au metallic surfaces and lower than that of polar MgO-MgO gaps. The difference in heat transfer coefficient is attributed to the presence of long-range electrostatic interactions in the $SiO_2$ structure, enhancing heat transfer compared to the Au-Au system. However, the strength of $SiO_2$ long-range Coulombic forces is smaller than MgO, which is confirmed by the difference in thermal conductivity between bulk MgO (~30-50 W/mK) and bulk $SiO_2$ (2-8 W/mK) [60,72].

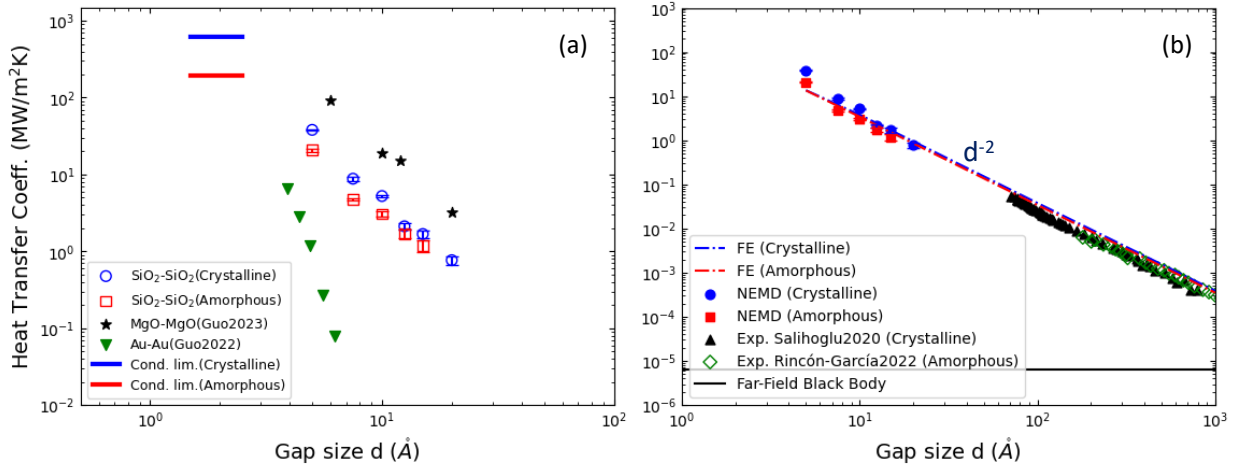

Fig. 2: (a) Heat transfer coefficient versus gap size for crystalline and amorphous silica (current study) and MgO-MgO [45] and Au-Au [28], all calculated by NEMD. The conduction limit is considered when two bodies are in contact (b) Fluctuational electrodynamics (FE) based on local dielectric function results for the heat transfer coefficient. Experimental data were obtained from Rincón-García et al. [54] and Salihoglu et al. [55].

The variation of the gap sizes from 0.5 nm to 2 nm results in a ~20-fold reduction in the heat transfer coefficient. Additionally, for all gap sizes, the heat transfer coefficient for the crystalline form of silica exceeds that of the amorphous form. Figure 2b presents the heat transfer coefficient between two silica plates using the fluctuational electrodynamics (FE) approach. The FE method utilizes the dielectric function extracted from the BKS potential, as discussed in the previous part of this work. Notably, there is a reasonable agreement between the results of NEMD and FE, with NEMD results following the $d^{-2}$ power trend. This alignment suggests that Coulombic electrostatic interaction facilitates heat transfer between the two slabs through optical vibrational modes, even in the presence of a gap. However, as the gap size exceeds 2 nm, It is difficult to compute heat transfer with molecular dynamics since the uncertainty due to numerical errors will be larger than the obtained values. It is important to note that for smaller gap sizes (~ 0.5 nm), molecular dynamics start to deviate from FE results, indicating the dominant role of acoustic vibrational modes (low-frequency phonons) as we will see.

Experimental results for gap sizes down to 7 nm for quartz silica by Salihoglu et al. [55] and down to 18 nm for amorphous silica by Garcia et al. [54] are also presented in Figure 2b. The NEMD results align with the experimental trend which shows a $d^{-2}$ power-law behavior. Furthermore, a comparison with black body radiation (continuous black line in Figure 2b) reveals that at a distance of 1 nm, the heat transfer coefficient between two semi-infinite silica surfaces is enhanced by a factor of 5-8 x $10^5$ compared to far-field black body radiation.

Figure 3 (a) to (f) show the spectral heat transfer for different gap sizes in both amorphous and crystalline silica structures. As is shown, the heat transfer coefficient for all frequencies decreases when the gap size increases. In addition, the peaks of the heat transfer coefficient occur at frequencies where the real part of the dielectric function (shown in Figure 2g,h) is minimum. The dielectric function has been calculated as detailed before by employing BKS potential. For the gap size of 0.5 nm, acoustic phonon tunneling becomes important and the heat transfer coefficient is enhanced in the low-frequency range (below 20 THz). This enhancement has no signature in the local dielectric function of the bulk material confirming the tunneling origin of this effect.

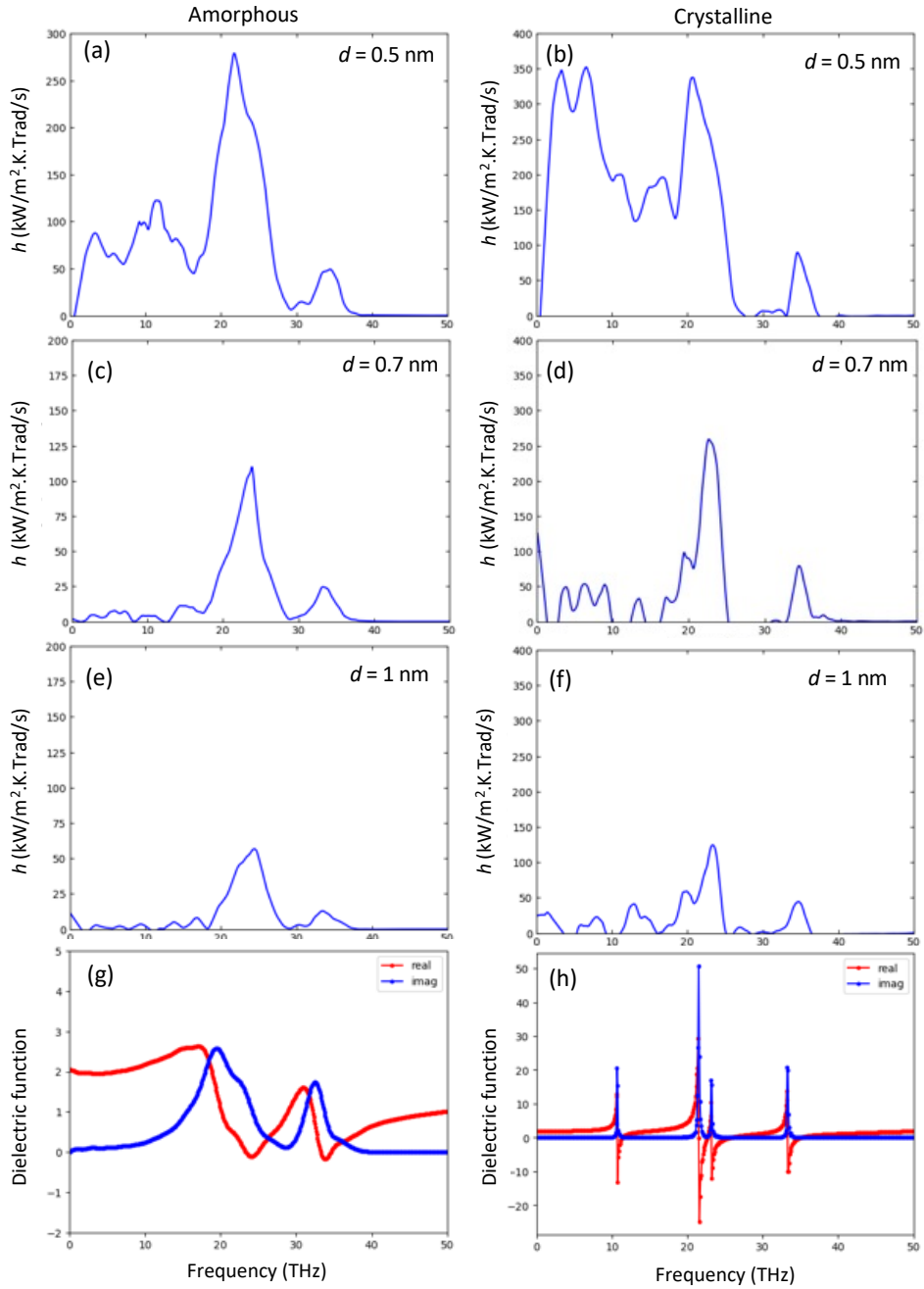

Fig. 3: (a)-(f): Spectral heat transfer coefficient versus frequency calculated by NEMD at different gap sizes for amorphous and crystalline $SiO_2$. (g), (h): The local dielectric function for bulk amorphous and crystalline $SiO_2$ computed by MD.

Fig. 4 shows the comparison of NEMD and FE (calculated based on the local dielectric function) for the heat transfer coefficient as a function of frequency for the two gap sizes of 0.5 nm and 1 nm. At the gap size of 1 nm, there is a good agreement between the two approaches particularly for the amorphous structure, while, for the gap size of 0.5 nm where acoustic phonons become important, the FE approach is not able to predict the contribution of low frequencies corresponding to tunneling acoustic modes.

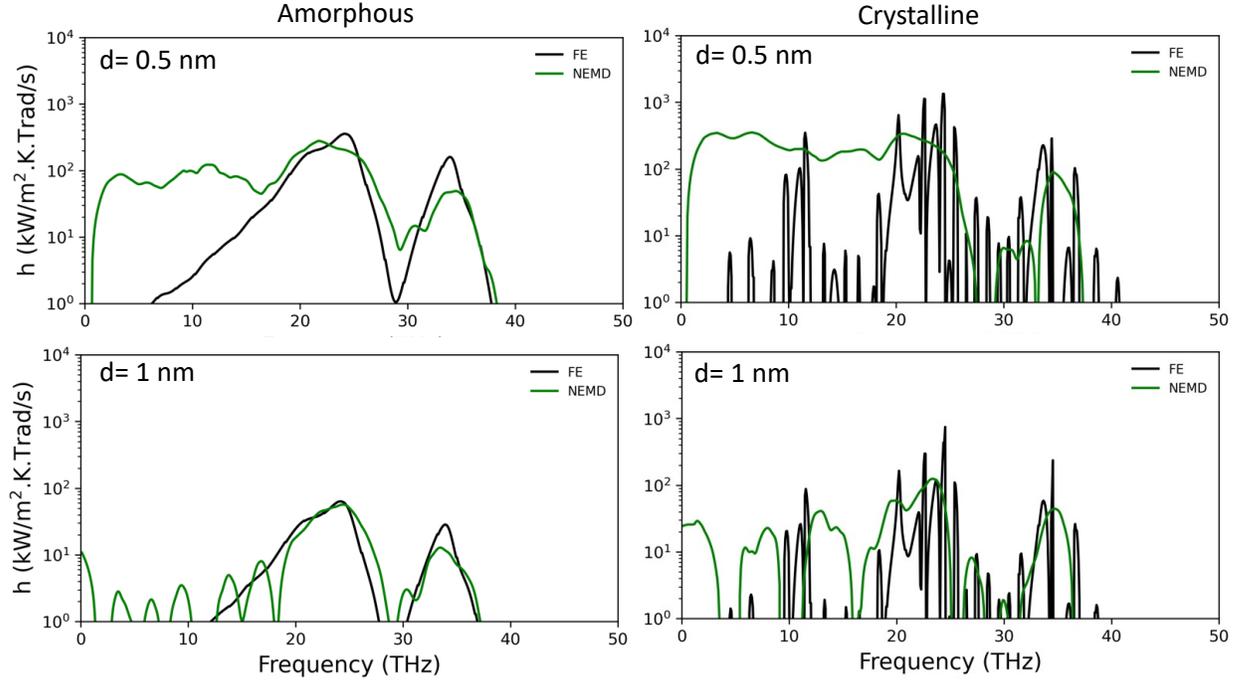

Fig. 4: Comparison between NEMD and FE (calculated based on local dielectric function) for spectral heat transfer coefficient at two gap sizes of 0.5 and 1 nm for both amorphous (left) and crystalline (right) forms of silica

As can be seen in Figure 4, for a gap size of 1 nm, the results of FE and NEMD for amorphous silica show an acceptable match, but they display differences in the crystalline silica. To better understand which phonon branches contribute to crystalline silica, we have calculated the nonlocal dielectric function of crystalline silica. For this purpose, the $\Gamma - X$ direction is considered in the direction of [100] and the wave vectors are considered as integer multiples of $2\pi/a_0$, with $a_0 = 0.5 \, nm$ that is very close the lattice parameter of crystalline silica. Figure 5(a) shows the contour plot of the imaginary part of the frequency- and wave-vector-dependent dielectric function in the $\Gamma - X$ direction for a bulk crystalline silica structure that is a cube with a length of 5 nm at the temperature of 300 K. By using the non-local dielectric function $\varepsilon(\omega, \boldsymbol{k})$, the heat transfer coefficient between two parallel plates can be calculated from Landauer's approach and fluctuational electrodynamics theory as follows [7,29,46,73]

$$h = \int_0^\infty \hbar\omega \, \partial f_{BE}/\partial T \, [\int_0^\infty \xi(\omega,\beta)\beta d\beta/2\pi ] \, d\omega/2\pi \tag{6}$$

where $\beta$ is the planar component of the wave vector, $f_{BE}$ is the Bose-Einstein distribution function, and $\xi$ is the energy transmission coefficient, dominated by the contribution of s and p-polarized evanescent waves, given by

$$\xi(\omega,\beta) \approx \sum_{y=s,p} \frac{4[\text{Im}(r_y)]^2 \exp[-2\,\text{Im}(k_z)d]}{|1-r_y^2 \exp[-2\,\text{Im}(k_z)d]|^2} \tag{7}$$

where $k_z = \sqrt{\left(\frac{\omega}{c}\right)^2 - \beta^2}$ and $r_s(\omega,\beta) = \frac{Z_s(\omega,\beta)-\frac{\omega}{k_z c^2}}{Z_s(\omega,\beta)+\frac{\omega}{k_z c^2}}$, $r_p(\omega,\beta) = \frac{\frac{k_z}{\omega}-Z_p(\omega,\beta)}{\frac{k_z}{\omega}+Z_p(\omega,\beta)}$ where $Z_s(\omega,\beta) = \frac{2j}{\pi\omega}\int_0^\infty \frac{dq_z}{\epsilon(\omega,k)-\left(\frac{ck}{\omega}\right)^2}$ and $Z_p(\omega,\beta) = \frac{2j}{\pi\omega}\int_0^\infty \frac{dq_z}{k^2}\left[\frac{q_z^2}{\epsilon(\omega,k)-\left(\frac{ck}{\omega}\right)^2} - \frac{\beta^2}{\epsilon(\omega,k)}\right]$

following the relation $k^2 = q_z^2 + \beta^2$, where we considered an isotropic dielectric function calculated from the $\Gamma - X$ direction.

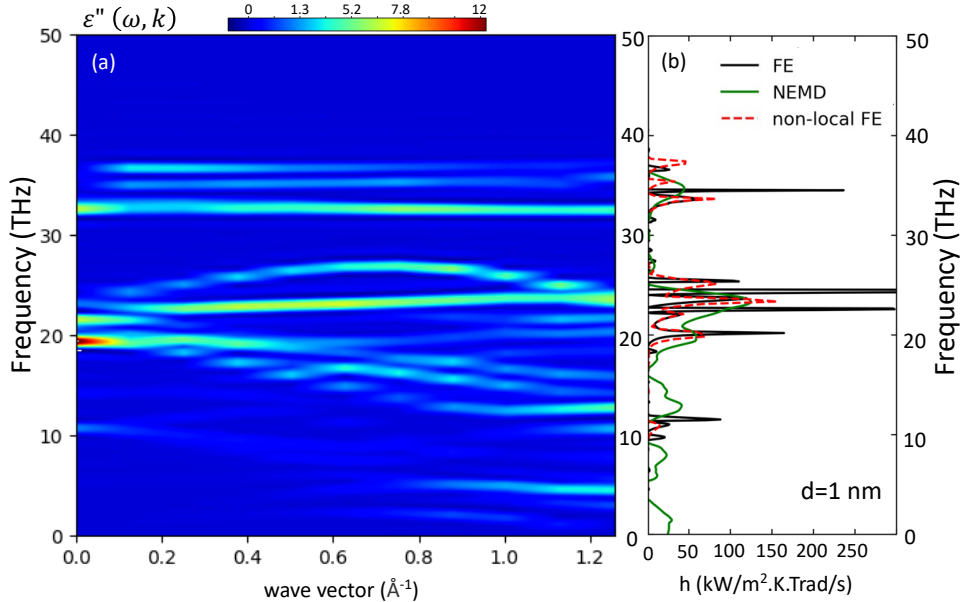

Fig.5: (a) Contour of the imaginary part of the dielectric function as a function of frequency and phonon wave vector for a bulk crystalline silica cube (b): Frequency-dependent heat transfer coefficient between two crystalline silica plates with a gap size of d=1 nm calculated by NEMD, FE and non-local FE.

Figure 5(b) shows the comparison of heat transfer coefficient results from NEMD, local FE, and non-local FE (dielectric function dependent on both wave vector and frequency) methods. As is shown, the results of the non-local approach predict the heat transfer coefficient in all frequency ranges, a lower peak than the local approach. While a better match is observed for non-local theory over its local counterpart, it's worth noting that neither approach, whether local or non-local, has yielded results entirely in line with molecular dynamics simulations. One reason for these relative discrepancies could be that the theory of fluctuational electrodynamics does not include the non-radiative acoustic phonon tunneling contribution to the heat exchange. Additionally, for the FE calculation, we used an isotropic dielectric function of bulk silica, which does not take into account the proximity of an interface. Therefore, we expect that if one calculates the spatially resolved dielectric function for two objects with a gap size of $d$ and employs it as input of the fluctuational electrodynamics approach, the results will be more consistent with the NEMD simulations.

In summary, we investigated the extreme near-field heat transfer between two silica surfaces, exploring gap distances below 2 nm through non-equilibrium molecular dynamics (NEMD) simulations. This investigation covered both amorphous and crystalline forms of silicon dioxide ($SiO_2$). Our findings revealed a substantial enhancement in heat transfer at sub-nanometric distances, showcasing the significance of both resonant surface waves and phonon tunneling. The NEMD simulations have been compared with fluctuational electrodynamics (FE) theory. The agreement between the heat transfer coefficient between $SiO_2$ planar objects from NEMD and FE theory, particularly with the $d^{-2}$ power-law behavior and alignment with experimental data above 7 nm and 18 nm for quartz and amorphous silica, respectively suggested that Coulombic electrostatic interactions play a crucial role in facilitating heat transfer through optical vibrational modes. Furthermore, spectral analyses revealed that the gap size significantly influences heat transfer, with peaks occurring at frequencies corresponding to minima in the real part of the dielectric function. Acoustic phonons, especially for small gap sizes, play a dominant role, explaining deviations between NEMD and FE predictions. The comparison of NEMD and FE approaches highlighted the importance of considering acoustic vibrational modes in the extreme near-field heat transfer regime, particularly for smaller gap sizes where tunneling effects become prominent. While observing reasonable agreement between NEMD and FE for amorphous silica at large gap sizes, considering non-local effects in the dielectric function for crystalline silica further refined the description, providing more accurate predictions for heat transfer coefficients in a broad range of frequencies. However, our atomic simulations call for the development of a theoretical framework that takes into account gradients of permittivity close to the material surfaces, going beyond the current non-local version of fluctuational electrodynamics. Our study contributes insights into the mechanisms governing extreme near-field heat transfer between silica surfaces, offering further exploration and application of these phenomena in emerging technologies.

**Acknowledgments**

This research received support from the French Agence Nationale de la Recherche (ANR) under Grant No. ANR-20-CE05-0021-01 (NearHeat). We also acknowledge the use of computational resources from Raptor at iLM, Université Claude Bernard Lyon 1.

**Data Availability**

The data that support the findings of this study are available from the corresponding authors upon reasonable request.

[45] Y. Guo, M. Gómez Viloria, R. Messina, P. Ben-Abdallah, and S. Merabia, Atomistic Modeling of Extreme Near-Field Heat Transport across Nanogaps between Two Polar Dielectric Materials, Phys. Rev. B 108, 085434 (2023).
[46] P.-O. Chapuis, S. Volz, C. Henkel, K. Joulain, and J.-J. Greffet, *Effects of Spatial Dispersion in Near-Field Radiative Heat Transfer between Two Parallel Metallic Surfaces*, Phys Rev B **77**, 035431 (2008).
[47] A. Fiorino, D. Thompson, L. Zhu, B. Song, P. Reddy, and E. Meyhofer, *Giant Enhancement in Radiative Heat Transfer in Sub-30 Nm Gaps of Plane Parallel Surfaces*, Nano Lett **18**, 3711 (2018).
[48] A. I. Volokitin and B. N. J. Persson, *Near-Field Radiative Heat Transfer between Closely Spaced Graphene and Amorphous SiO2*, Phys Rev B **83**, 241407(R) (2011).
[49] L. M. Zhang, G. O. Andreev, Z. Fei, A. S. McLeod, G. Dominguez, M. Thiemens, A. H. Castro-Neto, D. N. Basov, and M. M. Fogler, *Near-Field Spectroscopy of Silicon Dioxide Thin Films*, Phys Rev B **85**, 075419 (2012).
[50] G. Domingues, S. Volz, K. Joulain, and J.-J. Greffet, *Heat Transfer between Two Nanoparticles Through Near Field Interaction*, Phys Rev Lett **94**, 085901 (2005).
[51] S. Xiong, K. Yang, Y. A. Kosevich, Y. Chalopin, R. D'Agosta, P. Cortona, and S. Volz, *Classical to Quantum Transition of Heat Transfer between Two Silica Clusters*, Phys Rev Lett **112**, 114301 (2014).
[52] T. Ijiro and N. Yamada, *Near-Field Radiative Heat Transfer between Two Parallel SiO2 Plates with and without Microcavities*, Appl Phys Lett **106**, 023103 (2015).
[53] B. Song, D. Thompson, A. Fiorino, Y. Ganjeh, P. Reddy, and E. Meyhofer, *Radiative Heat Conductances between Dielectric and Metallic Parallel Plates with Nanoscale Gaps*, Nat Nanotechnol **11**, 509 (2016).
[54] L. Rincón-García, D. Thompson, R. Mittapally, N. Agraït, E. Meyhofer, and P. Reddy, *Enhancement and Saturation of Near-Field Radiative Heat Transfer in Nanogaps between Metallic Surfaces*, Phys Rev Lett **129**, 145901 (2022).
[55] H. Salihoglu, W. Nam, L. Traverso, M. Segovia, P. K. Venuthurumilli, W. Liu, Y. Wei, W. Li, and X. Xu, *Near-Field Thermal Radiation between Two Plates with Sub-10 Nm Vacuum Separation*, Nano Lett **20**, 6091 (2020).
[56] W. Humphrey, A. Dalke, and K. Schulten, *VMD: Visual Molecular Dynamics*, J Mol Graph **14**, 33 (1996).
[57] A. P. Thompson et al., *LAMMPS - a Flexible Simulation Tool for Particle-Based Materials Modeling at the Atomic, Meso, and Continuum Scales*, Comput Phys Commun **271**, 108171 (2022).
[58] B. W. H. van Beest, G. J. Kramer, and R. A. van Santen, *Force Fields for Silicas and Aluminophosphates Based on* Ab Initio *Calculations*, Phys Rev Lett **64**, 1955 (1990).
[59] G. J. Kramer, N. P. Farragher, B. W. H. van Beest, and R. A. van Santen, *Interatomic Force Fields for Silicas, Aluminophosphates, and Zeolites: Derivation Based on* Ab Initio *Calculations*, Phys Rev B **43**, 5068 (1991).
[60] A. J. H. McGaughey and M. Kaviany, *Thermal Conductivity Decomposition and Analysis Using Molecular Dynamics Simulations*, Int J Heat Mass Transf **47**, 1799 (2004).
[61] J. M. Larkin and A. J. H. McGaughey, *Thermal Conductivity Accumulation in Amorphous Silica and Amorphous Silicon*, Phys Rev B **89**, 144303 (2014).